\documentclass[prl,aps,twocolumn,amssymb]{revtex4} 
\usepackage{epsfig,amsmath}
\usepackage{psfrag}
\usepackage{subfigure}

\newcommand{\beq}{\begin{equation}}
\newcommand{\eeq}{\end{equation}}
\newcommand{\beqarr}{\begin{eqnarray}}
\newcommand{\eeqarr}{\end{eqnarray}}

\begin{document}

\title{Quasi regular concentric waves in heterogeneous lattices of coupled oscillators}

\author{Bernd Blasius and Ralf T{\"o}njes}
\affiliation{Institut f\"ur Physik, Universit\"at Potsdam,
Postfach 601553,D-14415 Potsdam, Germany}
\date{\today}

\begin{abstract}

We study the pattern formation in a lattice of coupled phase oscillators with
quenched disorder. In the synchronized regime concentric waves can arise, which
are induced and increase in regularity by the disorder of the system.  Maximal
regularity is found at the edge of the synchronization regime. The emergence of
the concentric waves is related to the symmetry breaking of the interaction
function. An explanation of the numerically observed phenomena is  given in a
one-dimensional chain of coupled phase oscillators. Scaling properties,
describing the target patterns are obtained.

\end{abstract}

\maketitle

The study of coupled oscillators is one of the fundamental problems in
theoretical physics and has led to many insights into the mechanisms of
spatio-temporal pattern formation in oscillatory media
\cite{Kuramoto85,Pikovsky01,Mikhailov04}, with applications in a variety of
systems such as arrays of Josephson junctions \cite{Wiesenfeld96} or oscillating
chemical reactions \cite{Kapral95}.  Renewed interest stems from its possible
role in many biological systems like cardiac tissue \cite{Winfree94}, neural
systems \cite{Varela01} and  ecological systems \cite{Blasius99}.  However,
nearly all theoretical studies have been carried out with idealized systems of
identical oscillators, whereas not much is known  about the dynamics and pattern
formation in heterogenous oscillatory media \cite{Kay00}. Here we show that
spatial disorder leads to the emergence of quasi regular concentric waves.

Target waves are one of the most prominent patterns in oscillatory media and are
usually associated with the presence of local impurities or defects in  the
system \cite{Zaikin70,Kuramoto85,Mikhailov94}.  These pacemakers change the
local oscillation frequency and are able to enslave all other oscillators in the
medium, which finally results in regular ring waves \cite{Kuramoto85,
Mikailov86}. However, the assumption of a discrete set of localized pacemaker
regions in an otherwise homogeneous medium is somewhat artificial. Especially
biological systems are often under the constraint of large  heterogeneity. In
such a disordered system no point can be distinguished as a pace-maker and it is
not clear whether such a system can sustain highly regular target patterns and
where they should originate.

In this paper we investigate the influence of quenched disorder on the pattern
formation in a lattice of  coupled phase oscillators.  As we show the random
nature of the medium itself plays a key role in the formation of the patterns.
As the disorder in a rather homogeneous synchronized medium is increased we
observe the formation of quasi regular target waves, which result from an
intricate interplay between the heterogeneity and a symmetry breaking of the
coupling function.

We study a system of $N$ coupled phase oscillators \cite{Kuramoto85}
\beq \dot{\theta}_i = \omega_i + 
\epsilon \sum_{j\in N_i} \Gamma(\theta_j-\theta_i), 
\quad i=1,\ldots , N.
\label{EqPhaseOsc} 
\eeq
Here, $\theta_i$ represents the phase of oscillator $i$, which is coupled with
strength $\epsilon$ to a set of nearest neighbors $N_i$ in a one- or
two-dimensional lattice. The natural frequencies $\omega_i$  are fixed in time,
uncorrelated and taken from a distribution $\rho(\omega)$. A scaling of time and
a transformation into a rotating reference frame can always  be applied so that
$\epsilon=1$ and the ensemble mean frequency $\overline{\omega}$ is equal to
zero. We refer to the variance $\sigma^2=\mathbf{var}(\omega_i)$ of the random
frequencies as the disorder of the medium.

The effects of coupling are represented by an interaction function $\Gamma$ 
which, in general, is a $2\pi$-periodic function of the phase difference. For
weakly coupled, weakly nonlinear oscillators  $\Gamma$ has the universal form
\cite{Kuramoto85,Koppel86}
\beq
\Gamma(\phi) = \left( \sin(\phi) + \gamma [1 - \cos (\phi)] \right)  .
\label{EqNonisoch}
\eeq
This function may be regarded as the first terms in a Fourier-expansion of
$\Gamma(\phi)$ with the constraint that $\Gamma(0)=0$.
The symmetry breaking  parameter $\gamma$ describes the nonisochronicity of the
oscillations \cite{Kuramoto85}.

\begin{figure}[tb]
(a)\includegraphics[width=3.2cm]{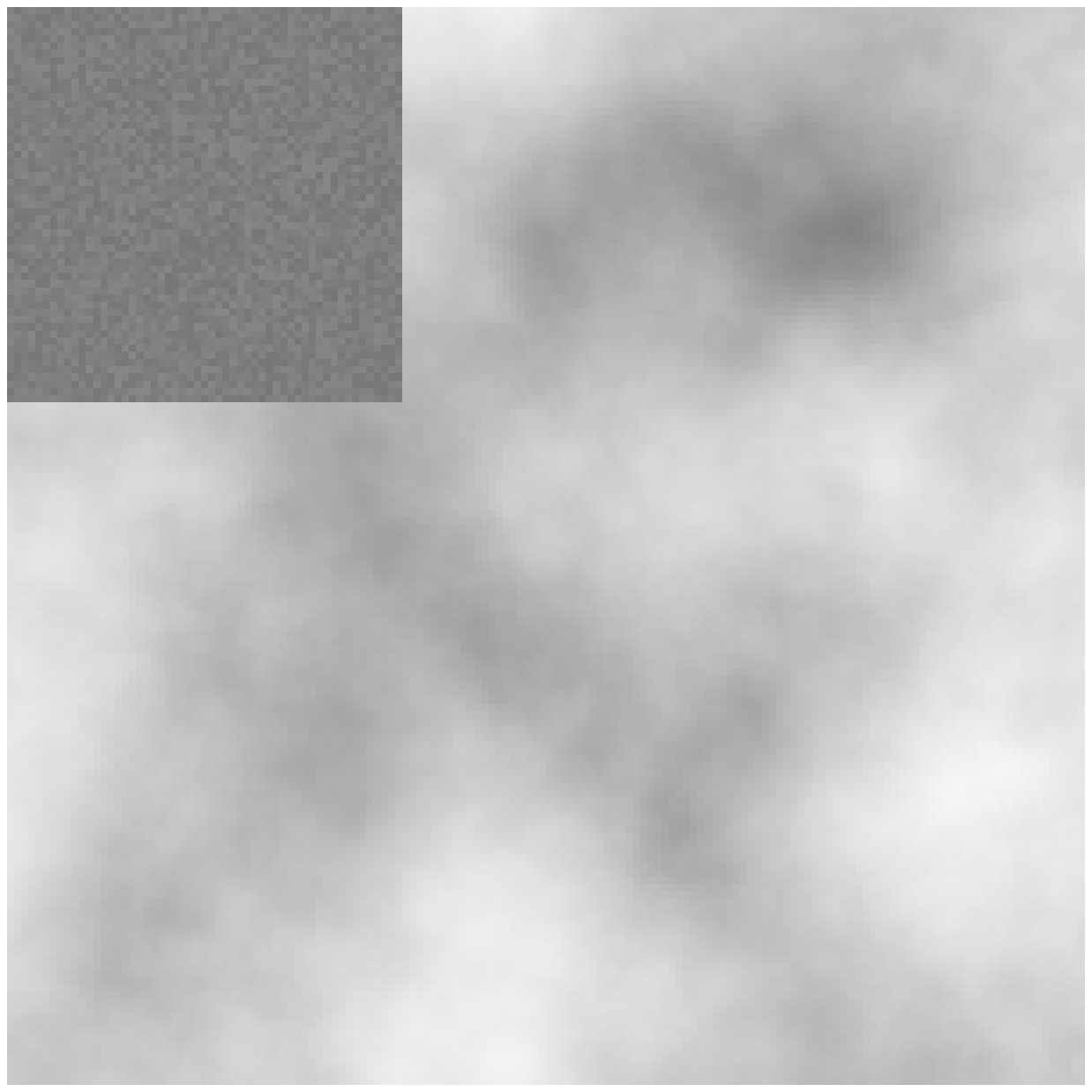}
(b)\includegraphics[width=3.2cm]{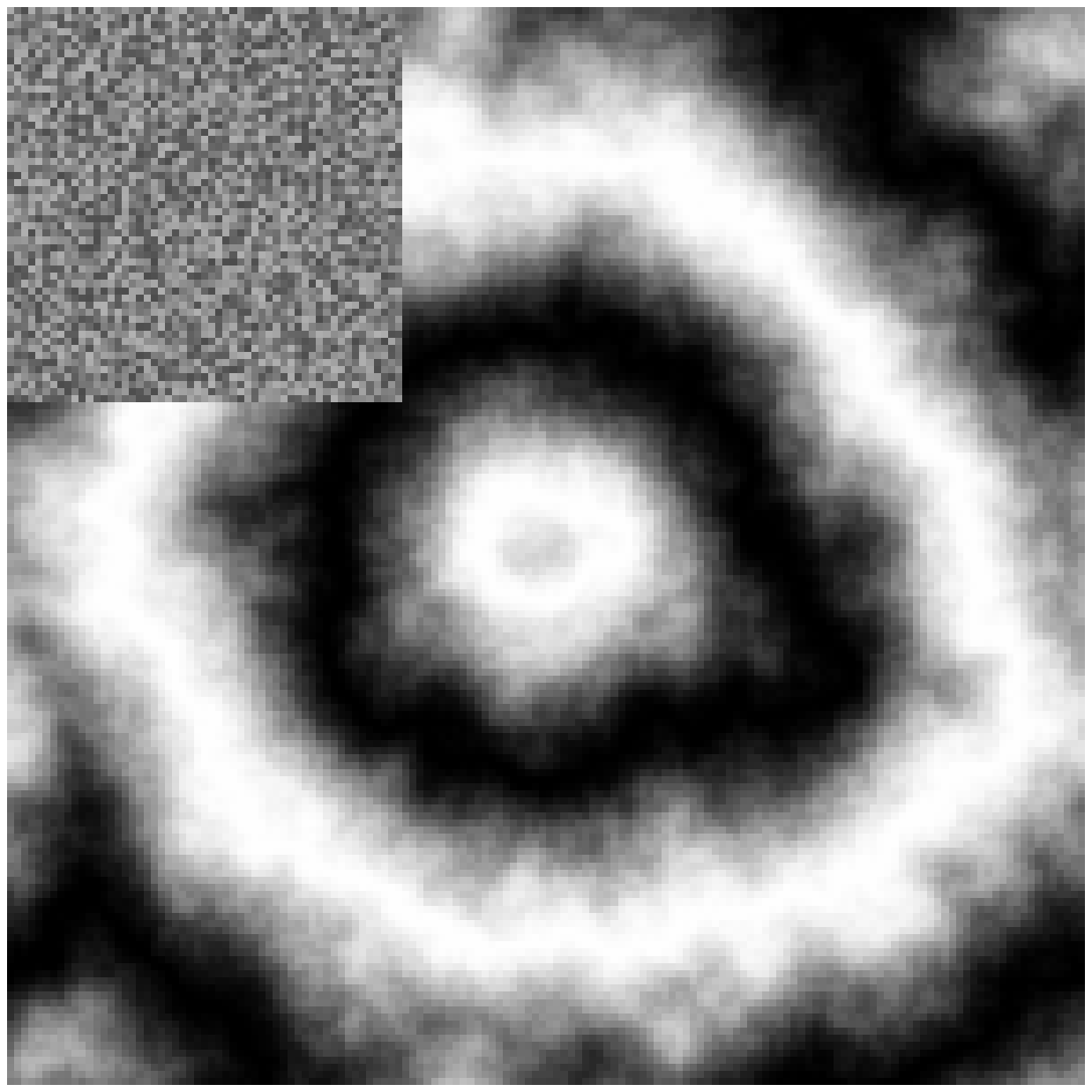} \\
(c)\includegraphics[width=3.2cm]{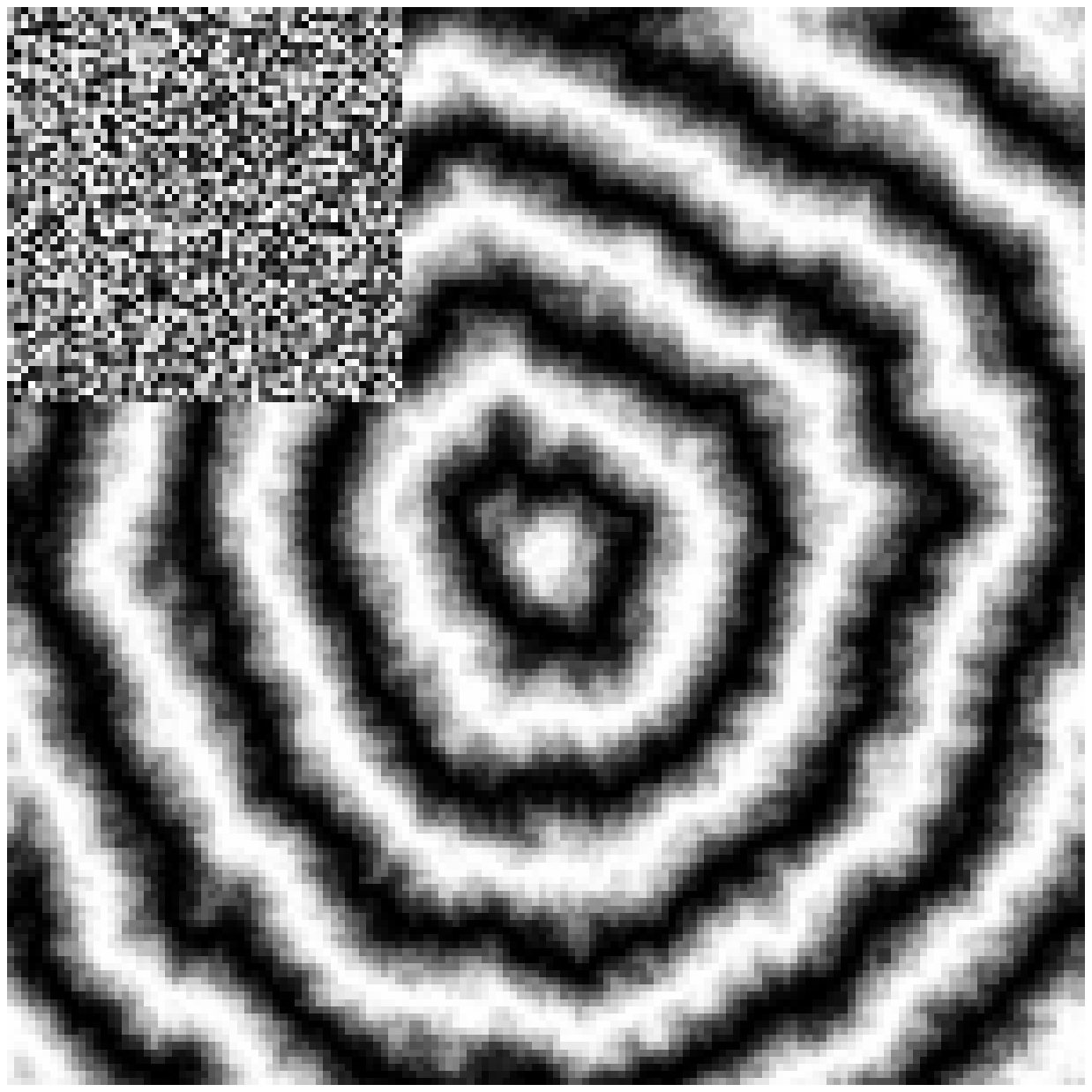} 
(d)\includegraphics[width=3.2cm]{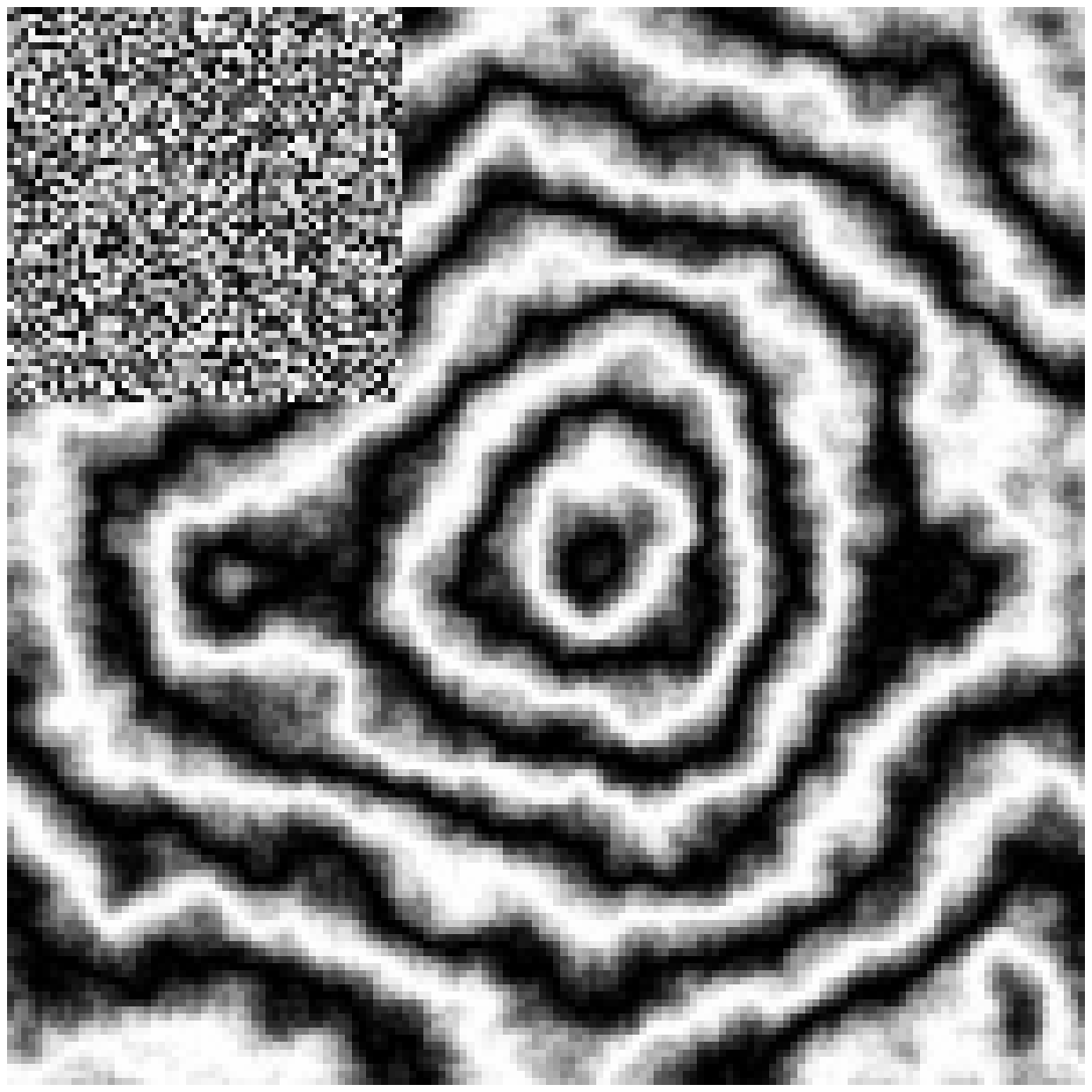} 
\caption{Simulation results in a 2-dimensional lattice of 150x150 phase
oscillators (\ref{EqPhaseOsc},\ref{EqNonisoch})  with nearest neighbor coupling,
$\epsilon=1$, periodic boundary conditions, homogeneous initial conditions and
(a) $\gamma=2$, $\sigma=0.029$, (b) $\gamma=2$, $\sigma=0.173$, (c) $\gamma=2$,
$\sigma=0.433$ and (d) $\gamma=0.3$, $\sigma=0.433$.  The random frequencies are
taken from a uniform distribution of variance $\sigma^2$.  Plotted is the sinus
of the phases $\theta_i$ as grey level. Similar results are obtained for open
boundaries. (a)-(c): effect of increased heterogeneity; (c)-(d): influence of
nonisochronicity $\gamma$. Insets show the natural frequencies $\omega_i$ as
grey levels.
}
\label{fig1}
\end{figure}

It is well known that for sufficiently small disorder the oscillators eventually
become entrained to a common locking frequency $\Omega$
\cite{Kuramoto85,Pikovsky01,Mikhailov04}. Since the oscillators are
nonidentical, even in this synchronized state they are usually separated by 
fixed  phase differences.  These  can sum up over the whole lattice to produce
spatio-temporal patterns, which are characterized by a stationary phase profile,
$\theta_i-\theta_1$.  This is demonstrated in Fig.~\ref{fig1}, where we have
simulated system (\ref{EqPhaseOsc})  with interaction (\ref{EqNonisoch}) in a
two-dimensional lattice. If the heterogeneity is small the oscillations across
the lattice synchronize to a homogenous phase profile (Fig.~\ref{fig1}a).
However, by increasing the disorder we observe the formation of  target waves
with decreasing wave length (Fig.~\ref{fig1}a-c).  The emergence of the
concentric waves is due to the symmetry breaking  in the interaction function.
For example, by reducing $\gamma$ in Eq.~(\ref{EqNonisoch}) the  pattern becomes more
irregular (Fig.~\ref{fig1}d).

It is the counterintuitive observation that the isotropic medium with random
frequency distributions of no spatial correlation (see Fig.~1 insets) can
generate and  sustain very regular wave patterns. Since the equations
(\ref{EqPhaseOsc}) generally approximate the phase dynamics for coupled limit
cycle oscillators this effect is not restricted to phase equations
(\ref{EqPhaseOsc}).  We have observed similar disorder induced target patterns
in lattices of a variety of oscillator types, including predator-prey and neural
systems, chemical reactions and even chaotic oscillators \cite{Blasius99}.

In the synchronized state all oscillators rotate with the constant locking
frequency $\dot{\theta}_i = \Omega$, so that system (\ref{EqPhaseOsc}) becomes a
set of $N$ equations, which have to be solved self consistently for the phases
$\theta_i$ and $\Omega$ under some imposed boundary conditions. To determine
$\Omega$ suppose first that the coupling function $\Gamma$ is fully
antisymmetric $\Gamma(-\phi)= -\Gamma(\phi)$, e.g. $\gamma=0$ in
Eq.(\ref{EqNonisoch}). In this case, by summing up all equations in
(\ref{EqPhaseOsc}) we obtain $\Omega = \overline{\omega}=0$ in the rotating
frame. Thus, nontrivial locking frequencies $\Omega\ne 0$  only arise if
$\Gamma(\phi)$ has a symmetric  part
$\Gamma_S(\phi)=\frac{1}{2}\left(\Gamma(\phi)+\Gamma(-\phi)\right)$,
\beq
\Omega = \frac{1}{N}\sum_{i,\\j\in N_i}\Gamma_S(\theta_j-\theta_i) .
\label{EqOmega}
\eeq

For any coupling function $\Gamma$ given  a realization of the  natural
frequencies $\omega_i$ we ask for the resulting phase profile $\theta_i$. Note,
that the inverse problem is easy to solve: for any regular phase profile
$\theta_i$ we can calculate $\Omega$ from Eq.~(\ref{EqOmega}), which after
inserting into Eq.(\ref{EqPhaseOsc}) yields the frequencies $\omega_i$.

\begin{figure}
\includegraphics[width=8.0cm,clip=true]{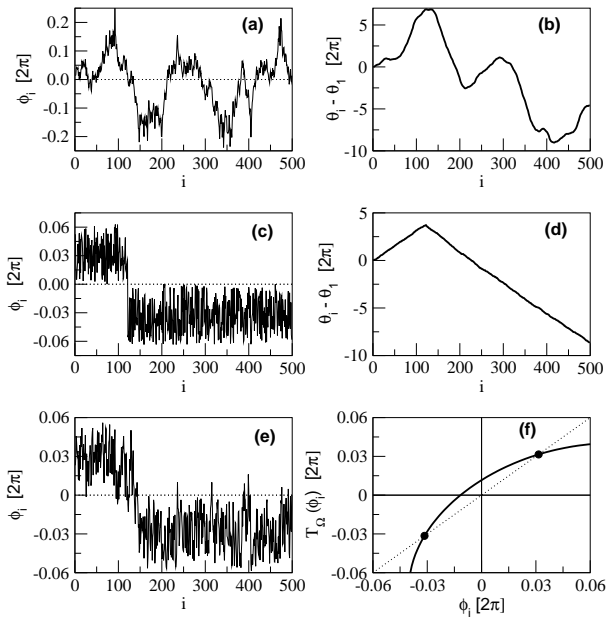} 
\caption{Phase differences $\phi_i$ (left) and phase profile $\theta_i -
\theta_1$ (b,d) in units of $2\pi$ for a chain of 500 phase oscillators
(\ref{EqPhaseChain}) with uniformly distributed frequencies
$\omega_i \in[0,0.2]$ and open boundaries.
(a,b) antisymmetric coupling (\ref{EqNonisoch}) with $\gamma=0$.
(c,d) unidirectional coupling (\ref{EqUnidir}) with $\Gamma(\phi)=\phi\,
\Theta(\phi)$.
(e) System (\ref{EqNonisoch}) with $\gamma=2$.
(f) Transfer map $T_\Omega (\phi_i)$ (\ref{EqTMap}) for Eqs. (\ref{EqNonisoch}) 
with $\gamma=2$ (solid line) and fixed points $\phi^*$ (filled circles).
}
\label{fig2}
\end{figure}

Insights into the pattern formation can be gained from a one-dimensional chain
of phase oscillators  \cite{Cohen82,Sakaguchi88}
\beq \Omega = \omega_i + \left[ \Gamma ( \phi_i ) + \Gamma (-\phi_{i-1}) \right].
\label{EqPhaseChain}
\eeq
Here we use $\phi_i=\theta_{i+1}-\theta_i$ for the phase differences between
neighboring oscillators. We assume open boundary conditions $\phi_0= \phi_N =
0$. The self consistency problem is trivial for an antisymmetric $\Gamma$  where
$\Omega=0$.   In this case the $\Gamma(\phi_i)$ simply describe a random walk
$\Gamma(\phi_i) = -\sum_{j=1}^i \omega_j$. Thus, for small $|\phi_i|$ the phase
profile $\theta_i$ essentially is given by a double summation, i.e. a
smoothening, over the disorder $\omega_i$ (see Fig.\ref{fig2}a,b).  Note, that
synchronization can only be achieved as long as the random walk  stays within
the range of $\Gamma$.  Thus, with increasing system size $N$ synchronization
becomes more and more unlikely.

The emergence of target waves is connected to a breaking of the coupling
symmetry. To explore this we study a unidirectional coupling with respect to
$\phi$
\beq \Gamma(\phi) = 
f(\phi) \, \Theta(\phi), \quad \mbox{for } |\phi| \ll 1,
 \label{EqUnidir}
\eeq
with the Heaviside function $\Theta(\phi)$ and  $f(\phi\! > \!0)>0$. Here, the
phase of oscillator $i$  is only influenced from neighboring oscillators which
are ahead of $i$. If the solution are small phase differences we are not
concerned about the periodicity of $\Gamma(\phi)$ as the coupling is only
required to be unidirectional close to zero.  For open boundaries
$\phi_0=\phi_N=0$ the solution to (\ref{EqPhaseChain},\ref{EqUnidir}) is given
by
\beq
\phi_i = \left\{
\begin{array}{ll} 
\quad  f^{-1}\left(\Omega-\omega_i\right)\, ,\quad  & i<m\\ 
-f^{-1}\left(\Omega-\omega_{i+1}\right)\, ,\quad  & i \ge m .
\end{array} \right.
\label{EqSoluUni}
\eeq
Here, the index $m$ is the location of the oscillator with the largest natural
frequency, which also sets the synchronization frequency 
\beq
\Omega = \omega_m = \mathbf{max}_i(\omega_i).
\label{EqMaxOm}         
\eeq
The phase differences (\ref{EqSoluUni}) are positive to the left of the fastest
oscillator, $i<m$, and negative to the right $i\ge m$. As a consequence, the
phase profile has a tent shape with a mean slope that is given by averaging
(\ref{EqSoluUni}) with respect to the frequency distribution (Fig.\ref{fig2}
c,d).   We call this solution type a quasi-regular concentric wave. This example
illustrates, that the asymmetry of the coupling function increases the influence
of faster oscillators and  effectively creates pacemakers with the potential to
entrain the whole system.   Note, that the solution
(\ref{EqSoluUni},\ref{EqMaxOm}) is not possible without disorder, i.e. for
$\sigma=0$.  Further, in contrast to the antisymmetric coupling, here
synchronization can be achieved for chains of arbitrary length.

In general, the coupling function $\Gamma$ will interpolate between the two
extremes  of fully antisymmetry and unidirectional coupling in the vicinity of
zero, e.g.  Eq.~(\ref{EqNonisoch}) with $\gamma\neq0$. As shown in Fig.2e  this
also gives rise to quasi regular concentric waves, very similar to the exactly
solvable system Fig.2c. To further investigate 
the origin of these patterns note that for
any given $\Omega$  system (\ref{EqPhaseChain})  implicitly defines two transfer
maps, $T_\Omega: \{\phi_{i-1},\omega_i\}\mapsto\phi_i$ and 
$T^{-1}_\Omega:\{\phi_i,\omega_i\}\mapsto\phi_{i-1}$, which describe the
evolution of the phase differences into the right or the left direction of the
chain, respectively. The random frequencies $\omega_i$ can be seen as noise
acting on the map (see Fig.2f)
\beq
\phi_i = T_\Omega(\phi_{i-1},\omega_i) =
\Gamma^{-1}\left[\Omega-\omega_i-\Gamma(-\phi_{i-1})\right] .
\label{EqTMap}
\eeq    
It is easy to see that the breaking of symmetry leads to a pair of fixed points,
$\phi^*$ and $-\phi^*$, in the noisy maps
\beq
\phi^* = T_\Omega(\phi^*,\overline{\omega}) = \Gamma_S^{-1}\left(\tfrac{\Omega}{2}\right).
\label{EqFixP}
\eeq
The transfer map can be linearized at the fixed points so that
$T_{\Omega}(\phi^*+\zeta,\omega)  \approx  \phi^* + a \zeta -  b \omega $ with
$a=\frac{\Gamma'(-\phi^*)}{\Gamma'(\phi^*)}$ and $b=\frac{1}{\Gamma'(\phi^*)}$. 
While one fixed point, $\phi^*$ in the case (\ref{EqNonisoch}) with $\gamma>0$, 
is linearly stable ($a\le 1$) the other fixed point is necessarily unstable.
These stability properties are inverted for $T_{\Omega}^{-1}$. Thus, when
iterating to the right of the chain the $\phi_i$  are concentrated around
$\phi^*$ and around $-\phi^*$ when iterating to the left. As a consequence, the
general solution of the selfconsistency problem (\ref{EqPhaseChain}) is build up
from two branches around the two fixed points  $\pm\phi^*$, superimposed by
autocorrelated fluctuations $\zeta_i$ (see Fig.\ref{fig2})
\beq
\phi_i =  \pm \phi^* + \zeta_i.
\label{EqGenSol}
\eeq
After summation this leads to the quasi regular tent shape of the  phase profile
$\theta_i$. In linear approximation the $\zeta_i$ describe  an AR(1) process
\beq
\zeta_i=a \zeta_{i-1} -  b \omega_{i}. 
\label{EqAR1}
\eeq
We want to stress that  the fluctuations $\zeta_i$ are an essential ingredient
of the solution. Although the emerging concentric waves seem to be regular the
underlying heterogeneity of the system does not permit analytical traveling wave
solutions  $\theta_i(t)=\Omega t - k |i-m|$.

The general solution (\ref{EqGenSol}) allows for very different phase profiles
(see Fig.~\ref{fig2}). The regularity of the wave pattern depends on the
relative influence of the mean slope $\phi^*$ compared to the fluctuations
$\zeta_i$ and can be measured by the quality factor
$Q_{\phi}={\phi^*}^2/\textbf{var}(|\phi_i|)$ and the autocorrelation $r$ of the
$\phi_i$. As demonstrated in Fig.\ref{fig3} both $Q_\phi$ and $r$ only depend on
the product $\gamma\sigma$ (see below). For $\gamma\sigma \to 0$ we find
$Q_{\phi}\to 0$ and $r\to 1$, and the solution is essentially a random walk (see
Fig.2a,b). With increasing values of $\gamma\sigma$ the correlations $r$ are
reduced and eventually become negative. Furthermore $Q_\phi$ increases with the
product $\gamma \sigma$, and for $\gamma>1$ can rise drastically (see
\ref{fig3}c). Thus, with increasing disorder of the system we obtain more
regular patterns until synchronization is lost.

A straightforward  integration of  system (\ref{EqPhaseOsc}) can be problematic
due to the long transients.  During the formation of the tent-profile initially
several locally synchronized clusters appear, with frequencies $\Omega_i$ that
are determined by the $\omega_i$ in the vicinity to the cluster centers. Upon
collision these clusters compete, where for $\gamma >0$ a higher-frequency
cluster will suppress a slower one \cite{Mikailov86}. The transient time for
extinction goes with $T \sim 1/\Delta \Omega$. In the disordered system   with
uniform $\rho(\omega)$ many local maxima with nearly identical frequencies
exist, so that the frequency differences $\Delta \Omega$ can become arbitrary
small with increasing system size.

Another approach, which also applies for two dimensional lattices, 
relies on the Cole-Hopf transformation of system (\ref{EqPhaseOsc}).
Assume that the $\phi_i$ are small so that  it is possible to approximate the
coupling  function (\ref{EqNonisoch}) around zero by  $\Gamma(\phi) =
\frac{1}{\gamma}\left(e^{\gamma\phi}-1\right) + O(\phi^3)$. After the Cole-Hopf
transformation $\theta_i=\frac{1}{\gamma}\ln q_i$ the synchronized lattice
(\ref{EqPhaseOsc}) is reduced to a linear system
\cite{Kuramoto85,Sakaguchi88,Mikhailov94}
\beq
\dot{q_i} = E q_i = \gamma \sigma \eta_i \cdot q_i + \sum_{j\in N_i} \left(q_j-q_i\right)
\label{EqSchr}
\eeq
where the random frequencies $\eta_i=\omega_i/\sigma$ are of zero mean and
variance one and $E=\gamma\Omega$ is some eigenvalue.  System (\ref{EqSchr}) is
known as the tight binding model for a particle in  a random potential on a
lattice \cite{Anderson58}. The eigenvector $\mathbf{q}^{\mbox{\tiny{max}}}$
corresponding to the largest eigenvalue $E_{max}$ will, in the re-transformed
system of angles, outgrow the contribution of all other eigenvectors to  the
time dependent solution linearly in time.
If the largest eigenvalue is non degenerate, the unique synchronized solution is
\beq
\theta_i(t)-\theta_1(t) =
\tfrac{1}{\gamma}\log\left(\tfrac{q^{\mbox{\tiny{max}}}_i}{q^{\mbox{\tiny{max}}}_1}\right).
\label{EqSchrSol}
\eeq
Eq.~(\ref{EqSchrSol}) is well defined since the components of
$\mathbf{q}^{\mbox{\tiny{max}}}$ do not change sign. Anderson localization
theory \cite{Anderson58} predicts exponentially decaying localized states with
some localization length \textit{l}, which after applying the reverse Cole-Hopf
transformation  yields the observed tent-shape phase profile with wavelength
$\lambda \sim \gamma \mbox{\textit{l}}$. Concentric waves emerge when $\lambda$
becomes smaller than the system size.

\begin{figure}
\includegraphics[width=8.5cm,clip=true]{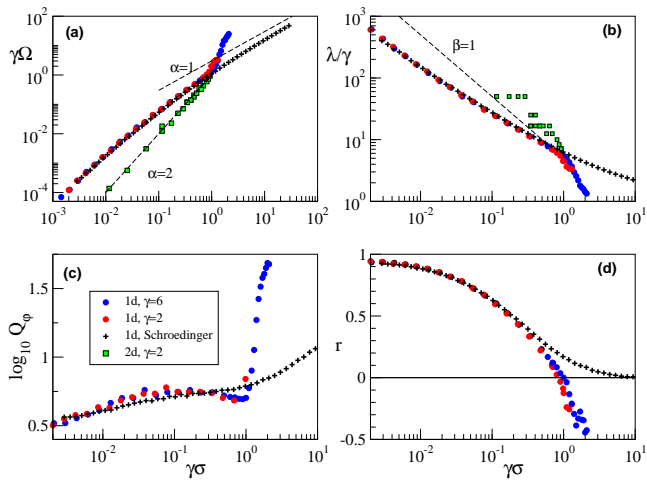} 
\caption{
Characterization of wave patterns by the locking frequency $\gamma\Omega$
(a),  the wave length $\lambda /\gamma$ (b),  the quality factor $Q_\phi$ (c)
and the cross correlation $r$ between neighboring phase differences (d). 
Numerical solutions are obtained by integrating system
(\ref{EqPhaseOsc},\ref{EqNonisoch}) for one dimensional (circles) and two
dimensional lattices (squares) in the synchronization regime. 
In one dimension the integrations were carried
out with chains of length 500 and averaged over 50 simulation runs 
for $\gamma=2$ (red circles) and $\gamma=6$ (blue circles). 
In the two dimensional system  each point (green squares) represents one single
simulation in a 100x100 lattice with $\gamma=2$.
The results using the eigenvector method (\ref{EqSchr}) are shown as (black +). Each point
represents an average of 500 simulations with $N=256$.
Further, indicated in (a) (b) are straight lines with a given
exponent $\alpha$ and $\beta$ (dashed lines).
The wavelength was obtained
for one dimension as $\lambda=2\pi/\phi^*$ and in the two dimensional
system  from a Fourier analysis of the phase profile. The plateaus in the
wavelength plot are finite size effects.}
\label{fig3}
\end{figure}

For extremal   values of $\gamma\sigma$  the system (\ref{EqSchr}) has well
defined scaling properties $E_{max} \sim \left(\gamma\sigma\right)^{\alpha}$ and
$\mbox{\textit{l}}\sim \left(\gamma\sigma\right)^{-\beta}$
\cite{Anderson58}.  Perturbation theory
yields $\alpha=2$ for $\gamma\sigma\ll 1$ and $\alpha=1$ for $\gamma\sigma\gg
1$.  For the exponent $\beta$ we find $\beta \lesssim 1$ in the one dimensional
system, while $1 \le \beta \le 2$ in the two dimensional lattice.   This implies
for the synchronization frequency $\Omega$ and the wavelength $\lambda$
\beq
\Omega  \sim \gamma^{\alpha-1}\sigma^{\alpha}, \qquad
\lambda  \sim \gamma^{1-\beta}\sigma^{-\beta} .
\label{EqOhmLambScale}
\eeq
Here, $\gamma$ does not influence the wavelength as  much as $\sigma$ but while an
increase of $\gamma$ in one dimension leads also to an increasing wavelength the
effect in two dimensions is the opposite.

In one dimension the scaling with $\gamma\sigma$ holds as long as Eq.
(\ref{EqSchr}) can approximate the transfer map reasonably. The approximation
breaks down when  the correlation $r$ of the phase differences becomes negative.
In this regime the quality factor $Q_\phi$ strongly depends on both
$\gamma\sigma$ and $\gamma$. The noise term  $b \omega_{i}$ in (\ref{EqAR1}) can
become very small with increasing $\gamma$. This regime, which is not described
by the Anderson approximation, can produce very regular concentric waves near
the border of desynchronization.

The constructive role of noise has often been studied \cite{Shinbrot01}. It has
been shown that in spatially extended excitable systems noise can enhance the pattern
formation and for example is able to promote traveling waves
\cite{Kadar98,Jung95}. Further, it is well known that disordered systems can
synchronize faster \cite{Braiman95}.  Here, we have investigated the influence of
quenched noise on the pattern formation in the oscillatory regime. Whereas local coupling tends to
synchronize the oscillators, the imposed disorder tends to desynchronize the
array. We have shown that the tension between these two opposing forces can lead
to concentric target patterns.

This work was supported by the German Volkswagen Stiftung and SFB 555.

\end{document}